\documentclass[aps,twocolumn,superscriptaddress,10pt]{revtex4-1}

\usepackage{algcompatible}
\usepackage{amsmath}
\usepackage{amssymb}
\usepackage{bm}
\usepackage{newtxtext}
\usepackage{newtxmath}
\usepackage{booktabs}
\usepackage{graphicx}
\usepackage{newfloat}
\usepackage{float}
\usepackage{ctexheading}
\usepackage{hyperref}
\DeclareFloatingEnvironment[
  name=ALGORITHM,
  placement=tbhp,
]{algorithm}
\hypersetup{colorlinks=true}

\ctexset{
  section = {
    numbering = false,
    format = \Large\raggedright,
    afterskip = 1ex,
  },
  paragraph/numbering = false,
}

\newcommand{\Rom}[1]{\MakeUppercase{\romannumeral #1}}

\begin{document}

\title{Hard instance learning for quantum adiabatic prime factorization}
\author{Jian Lin}
\affiliation{State Key Laboratory of Surface Physics, Institute of Nanoelectronics and Quantum Computing, and Department of Physics, Fudan University, Shanghai 200433, China}

\author{Zhengfeng Zhang}
\affiliation{Shanghai Key Lab of Intelligent Information Processing,and School of Computer Science, Fudan University, Shanghai 200433, China}

\author{Junping Zhang}
\email{jpzhang@fudan.edu.cn}
\affiliation{Shanghai Key Lab of Intelligent Information Processing,and School of Computer Science, Fudan University, Shanghai 200433, China}

\author{Xiaopeng Li}
\email{xiaopeng\_li@fudan.edu.cn}
\affiliation{State Key Laboratory of Surface Physics, Institute of Nanoelectronics and Quantum Computing, and Department of Physics, Fudan University, Shanghai 200433, China}
\affiliation{Shanghai Qi Zhi Institute, AI Tower, Xuhui District, Shanghai 200232, China}

\begin{abstract}

\bfseries

Prime factorization is a difficult problem with classical computing, whose exponential hardness is the foundation of Rivest-Shamir-Adleman (RSA) cryptography. With programmable quantum devices, adiabatic quantum computing has been proposed as a plausible approach to solve prime factorization, having promising advantage over classical computing.
Here, we find there are certain hard instances that are consistently intractable for both classical simulated annealing and un-configured adiabatic quantum computing (AQC).
Aiming at an automated architecture for optimal configuration of quantum adiabatic factorization, we apply a deep reinforcement learning (RL) method to configure the AQC algorithm.
By setting the success probability of the worst-case problem instances as the reward to RL, we show the  AQC performance on the hard instances is dramatically improved by RL configuration. The success probability also becomes more evenly distributed over different problem instances, meaning the configured AQC is more stable as compared to the un-configured case.
Through a technique of transfer learning, we find prominent evidence that the framework of AQC configuration is scalable---the configured AQC as trained on five qubits remains working efficiently on nine qubits with a minimal amount of additional training cost.

\end{abstract}

\date{\today}


\maketitle


Prime factorization plays a vital role in information security as its computation complexity on classical computers forms the foundation of RSA cryptography. The capability of factorizing $N$ into a product of two prime integers $N = p\times q$ is enough to break the RSA cryptosystem. It has been attracting tremendous efforts in a broad range of sciences from heuristic algorithm design~\cite{gendreau2010handbook} and machine learning~\cite{meletiou2002first,Stekel_2018}, to bio-inspired computation~\cite{yampolskiy2010application,Monaco_2017} and stochastic architectures~\cite{borders2019integer}. Although this problem is not expected to be NP-hard, all the established classical algorithms have an exponential time cost in the size  of $\log N$,  by which RSA cryptosystem is secure. With quantum computing resources, Shor's quantum-circuit-based algorithm reduces the computation time cost to polynomial~\cite{shor1999polynomial}. However, in the present era of noisy intermediate size quantum (NISQ) technology~\cite{PreskillNISQ}, the experimental implementation of Shor's quantum factorization is largely restricted to small integers~\cite{vandersypen2001experimental,dash2018exact} due to its demanding requirement on the qubit number and gate quality.

An alternative approach to perform prime factorization on quantum devices is through adiabatic quantum computing (AQC)~\cite{farhi2001quantum,peng2008quantum}, where the factorization problem  is encoded into the ground state of a spin Hamiltonian $\mathbf{H}_{\text{P}}$. To start,  the quantum system is prepared in the ground state of a trivial Hamiltonian $\mathbf{H}_{\text{B}}$, and then let evolve under a time ($\tau$) dependent Hamiltonian
\begin{equation}
\mathbf{H} (\tau)= [1-\lambda(\tau/T)] \mathbf{H}_{\text{B}} + \lambda(\tau/T)\mathbf{H}_{\text{P}},
\label{eq:Ham}
\end{equation}
where $\lambda(\tau/T)$ is Hamiltonian schedule having $\lambda(0) = 0$ and $\lambda(1) = 1$, and $T$ is the total quantum evolution time, or equivalently the AQC computation time.

\begin{figure*}[htp]
\includegraphics[width=\textwidth]{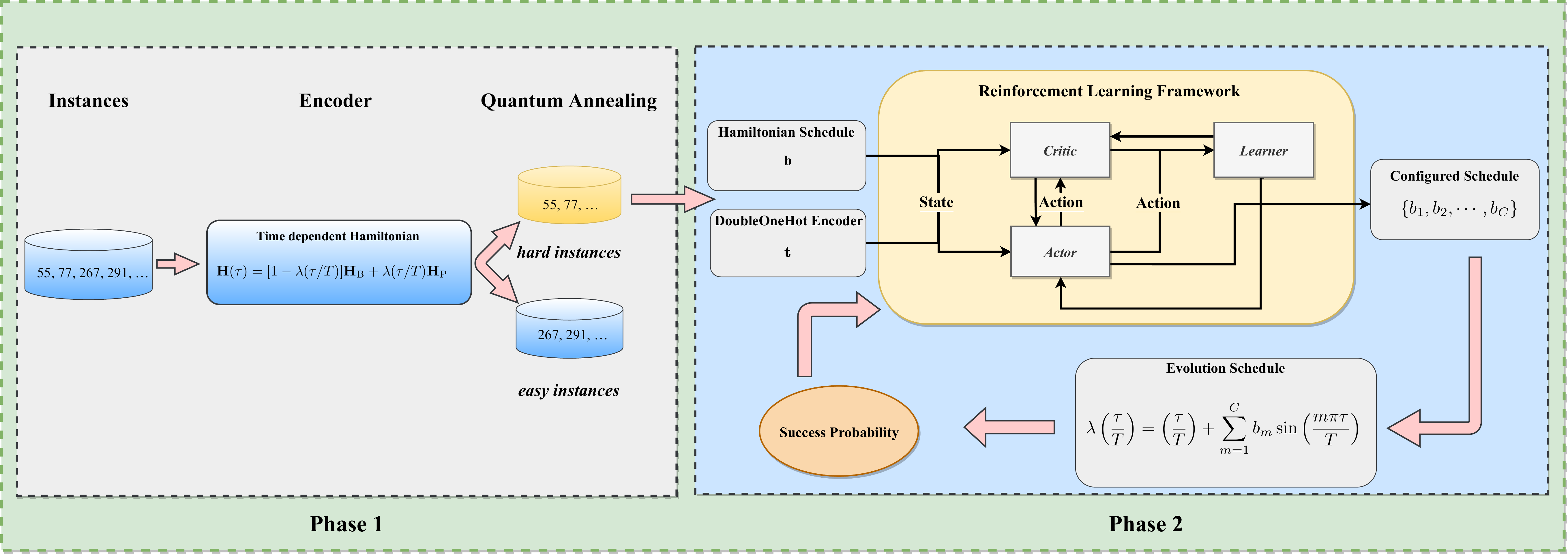}
\caption{\textbf{Schematic illustration of hard instance learning architecture for quantum adiabatic prime factorization.} In phase 1, we map the factorization problem to the quadratic unconstrained binary optimization (QUBO) problem and transfer it into an equivalent Ising type Hamiltonian (Supplementary Material). We generate instances that need the same number of qubits to factorize. Then, we separate factorization instances into two groups according to their performance in AQC and load the intractable  instances under the un-configured AQC schedule into the RL optimization process. In phase 2, we show the RL structure. We combine the Hamiltonian schedule (all the $b_m$ in Eq.~\eqref{eq:Schedule}) and DoubleOneHot Encoder as the input state to the neural networks. The critic neural network supervises the actor neural network to take actions on the Hamiltonian schedule. A quantum adiabatic computer works under the configured schedule and provides the success probability as feedback to the RL framework.}
\label{fig:architecture}
\end{figure*}

The AQC model has been used to solve prime factorization by taking a cost function  $(N-p\times q)^2$, with $p$ and $q$ in the binary representation and using a fixed Hamiltonian schedule as $\lambda(\tau/T) =(\tau/T)^2$~\cite{peng2008quantum}. The classical binary-formed cost function is directly promoted to a quantum Hamiltonian $\mathbf{H}_{\text{P}}$ using the quantum computation basis. One problem with this Hamiltonian encoding is the coupling strengths scale exponentially with $\log N$, which is unphysical. This problem is resolved with an improved encoding protocol incorporating the multiplication table~\cite{xu2011quantum,Jiang2018QuantumAF}. This approach has received much attention in recent years~\cite{dridi2017prime,xu2017experimental,albash2018adiabatic,wang2020prime}, as triggered by the fascinating progress achieved in quantum annealing devices~\cite{hauke2020perspectives,Deutsch_2020}. At the same time, concerns have been raised that the spin glass problem arising in the spin Hamiltonian encoding  may prohibit advantage in the AQC based prime factorization over classical solvers~\cite{mosca2019factoring,hauke2020perspectives}.
Here, we propose a scheme for AQC algorithm configuration based on reinforcement learning and apply it to prime factorization (see Figure~\ref{fig:architecture} for illustration). In the learning process, we use a reward setting that reflects the AQC performance on the most difficult factorization instances. Using a soft-actor-critic RL method, we find the learning process has an astonishing convergence speed---it converges within only a few hundred measurement steps, significantly faster as compared to previous studies using RL for quantum state preparation~\cite{bukov2018reinforcement}, for parameter configuration in quantum approximate optimization~\cite{wauters2020reinforcement}, and for adiabatic quantum algorithm design~\cite{lin2020quantum}, which takes about $10^4$ to $10^6$ measurement steps. The configured AQC algorithm produces an improved success probability, more evenly distributed over different factorization instances as compared with the un-configured algorithm. Through the numerical test, we show the approach of RL-based AQC algorithm configuration has fair transferability.

\begin{figure}[htp]
\includegraphics[width=.48\textwidth,height=12.5cm]{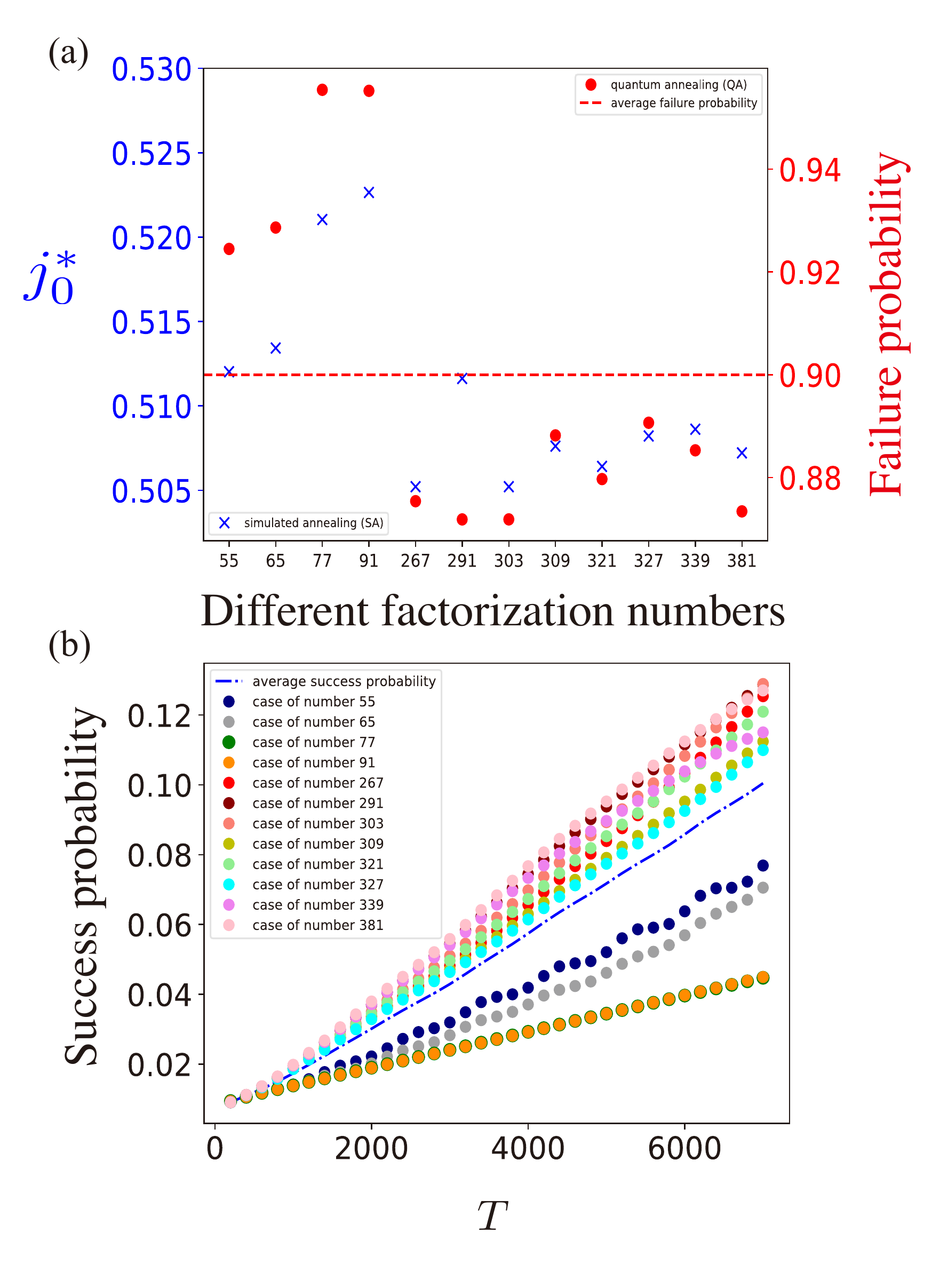}
\caption{\textbf{Easy and hard factorization instances with simulated annealing and adiabatic quantum computing.} We separate the problem instances into two groups according to their performances. In (a), blue data show the $j_0^*$ obtained in 500 parallel runs of simulated annealing (see main text). Red data show the failure probability of un-configured AQC (see main text) with a total evolution time $T = 6964$. The red dashed line denotes the average failure probability of all the problem instances. It reaches the threshold of failure probability which we set to be 0.9. The problem instances that are hard with SA which have a high value of $j_0^*$ remain hard for the un-configured AQC as it yields a considerably higher failure probability compared to other problem instances. In (b), we show the performances of different instances with the increasing total evolution time $T$. We find the success probabilities of several instances increase more slowly than others and are under the average value at the end. We separate them into two groups with the threshold (see main text). In these plots, we choose numbers to factorize that are encoded by 7 spins.
}
\label{fig:QASA}
\end{figure}

\section{Results}

\paragraph{Easy and hard factorization instances.}
We map the prime factorization problem into the quadratic unconstrained binary optimization (QUBO) problem~\cite{boros2007local,zeng2016schedule,lewis2017quadratic,patton2019efficiently} which can be directly transferred into Ising type Hamiltonian ($\mathbf{H}_{\text{P}}$)~\cite{Jiang2018QuantumAF,peng2019factoring,wang2020prime}.
The details are provided in the Supplementary Material. From the perspective of quantum spin glass, the major difficulty in reaching the ground state of a spin Hamiltonian is the existence of metastable spin configuration having large energy barriers~\cite{bapst2013quantum,mosca2019factoring}. Since the energy barriers cause slow relaxation in simulated annealing (SA) and make the algorithm inefficient, we group the different problem instances according to the efficiency of SA in reaching the true ground states.

Specifically, we take a classical spin system with energy defined by $\mathbf{H}_{\text{P}}/\left\|\mathbf{H}_{\text{P}}\right\|_\infty$, where
$\left\|\mathbf{H}_{\text{P}}\right\|_\infty$ denotes the maximum coefficient in QUBO type cost function of all the instances in the same system size, and perform the following SA protocol. Starting from a random spin configuration corresponding to an infinite temperature ensemble, the spin configuration is locally updated at a randomly picked site, and then accepted with a probability
 $P(j)=e^{-\beta(j) \Delta E}$ at each step (labeled by $j$), where $\Delta E$ is the energy cost of the local spin update.
 The inverse temperature increases step-by-step according to $\beta(j) = \beta_0e^{(j/j_0)}, j\in[0,10j_0]$. With a given $j_0$, there is a corresponding success probability to reach the actual ground state at the end of SA. The success probability tends to increase with $j_0$, since a slower SA has a better chance to succeed.
For each problem instance, we perform 500 parallel runs of SA. In each run, we increase $j_0$ until  the actual ground state is reached. The corresponding $j_0$ is defined to be $j_0^*$. This quantity measures the number of steps for SA to succeed and can thus be used to quantify  the energy barrier of the encoding spin glass problem and consequently the hardness of the computation.
We obtain the statistics of $j_0^*$ out of the parallel SA runs.
In Figure~\ref{fig:QASA}(a), we show the mean values of $j_0^*$ obtained in performing SA on factorizing 55, 65, 77, 91, 267, 291, 303, 309, 321, 327, 339, 381, which involves 7 bits in our Hamiltonian encoding. The numbers 77, 91 are found to be difficult to factorize, as the associated $j_0^*$ is typically larger than other numbers.

We also test the different performances of problem instances with AQC. The time evolution of the quantum system is formally described by a time-dependent quantum state
\begin{equation}
|\psi (\tau) \rangle  = {\cal T}_\tau e^{-i \int _0 ^\tau d\tau' \mathbf{H}(\tau') } |\psi(0) \rangle,
\label{eq:psit}
\end{equation}
where ${\cal T}_\tau$ represents time-ordering. The computation results of AQC are collected by collapsing the final quantum state in the computation basis. The results are stochastic in general. The success probability of AQC to collapse onto the correct solution tends to increase as we increase the adiabatic evolution time ($T$)~\cite{albash2018adiabatic}.
In the numerical test, we increase $T$ until the average success probability reaches a threshold $P_\mathrm{th}$, which is set at 0.1.
We separate the factorization instances into easy and hard groups according to their individual AQC success probabilities.
The problem instances having a success probability larger (smaller) than the  threshold  ($P_\mathrm{th}$) are counted as easy (hard).


From Eq.~\eqref{eq:psit}, the final quantum state remains intact as we rescale $\tau \to \zeta \tau$, and $\mathbf{H} \to \mathbf{H}/\zeta$, with $\zeta$ an arbitrary scaling factor.
For physical consideration, a Hamiltonian convention is imposed by,
\begin{equation}
\mathbf{H} (\tau) \to \mathbf{H} (\tau)/\left\|\mathbf{H}_{\text{P}}\right\|_\infty,
\end{equation}
where
$\left\|\mathbf{H}_{\text{P}}\right\|_\infty$ denotes the same meaning in simulated annealing. With this convention, the required interaction strength in a physical system is guaranteed to be bounded.

We have checked the performance of the un-configured AQC with a previously used Hamiltonian schedule $\lambda (\tau/T) = (\tau/T)^2$~\cite{peng2008quantum}.
As shown in Figure~\ref{fig:QASA}(a), there are several
factorization instances whose failure probabilities are substantially larger than other typical instances,
e.g., with  $N = 77, 91$.
As shown in Figure~\ref{fig:QASA}(b),  the success probabilities of several cases increase quite slowly with $T$. The factorization problem instances that are hard with SA appear to remain hard on the un-configured AQC.

Aiming at rescuing the hard factorization instances, we develop an RL-based configuration scheme for  the AQC algorithm (see Figure~\ref{fig:architecture} for illustration). It is known in quantum adiabatic Grover search that different choices of Hamiltonian schedule lead to different computation complexity~\cite{roland2002quantum,lin2020quantum}. There are different choices for configuring the AQC algorithm in principle.  Here we choose to vary the standard Hamiltonian schedule $\lambda (\tau/T)$ to improve the algorithm performance. 
We parameterize the Hamiltonian schedule as,
\begin{equation}
\lambda \left(\frac{\tau}{T}\right) = \left(\frac{\tau }{T} \right) + \sum_ {m=1} ^{C} b _m  \sin\left(\frac{m\pi \tau}{T}\right).
\label{eq:Schedule}
\end{equation}
Here the schedule satisfies the boundary conditions $\lambda(0)  = 0$ and $\lambda(1) = 1$, and the parametrization is complete if we take the high frequency cutoff $C \to \infty$. The parameters ($b_m$) are denoted by a vector $\mathbf{b}$ in the following.
With $\mathbf{b} \to 0$, the schedule goes back to the linear form as used in standard AQC~\cite{albash2018adiabatic,xu2017experimental,hauke2020perspectives,qiu2020programmable,saxena2021hybrid}.

\begin{figure}[htp]
\includegraphics[width=.48\textwidth,height=7cm]{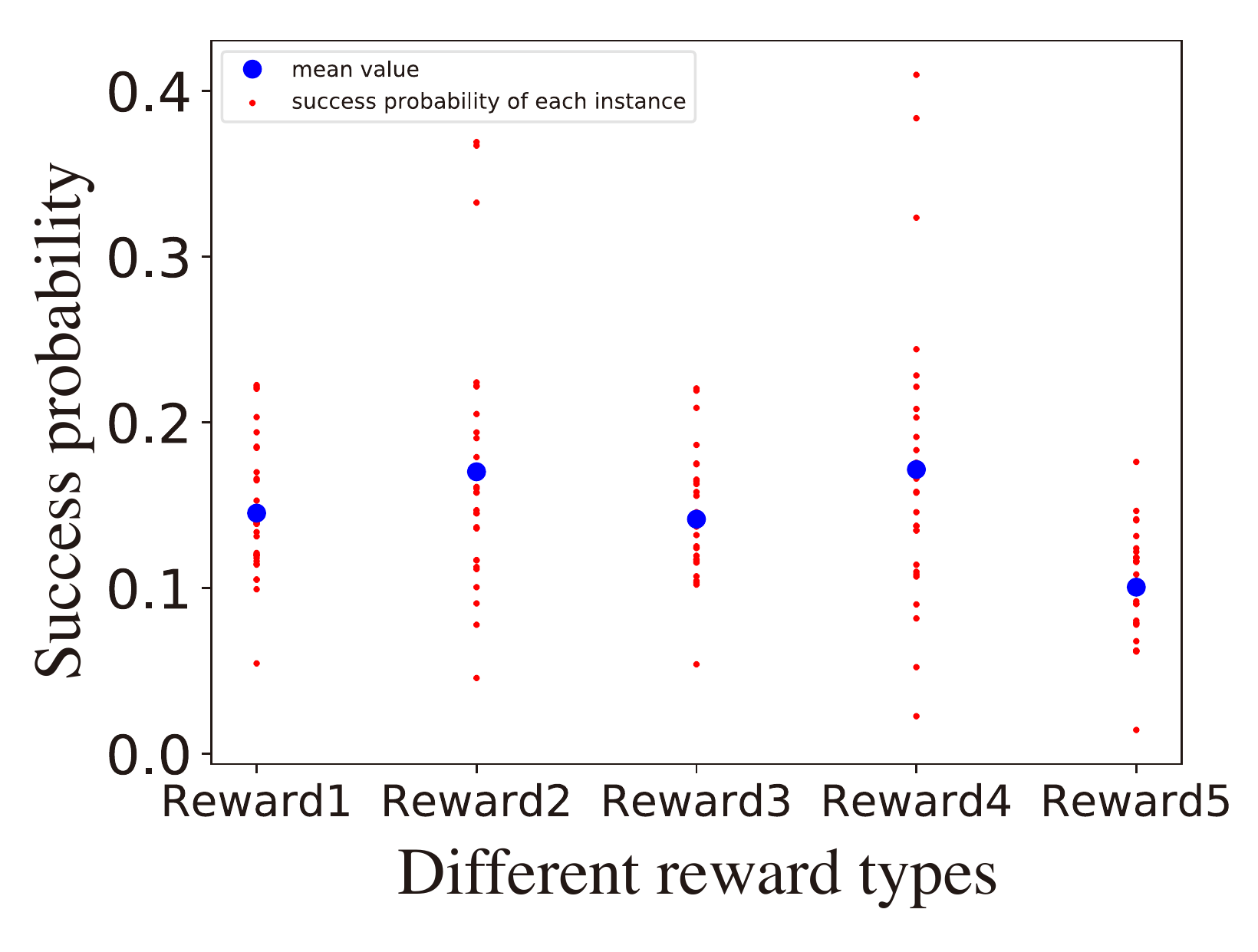}

\caption{ \textbf{The performance of SAC configured AQC with different reward settings.} The success probabilities on   all the hard factorization instances are obtained by using the best Hamiltonian schedule that is trained under the different reward settings within the same number of training steps. We test the different reward types including Reward1: {\it $\min$($\log$(success probability))}, Reward2: \text{ave}{\it ($\log$(success probability))}, Reward3: {\it $\min$(success probability)}, Reward4: \text{ave}{\it (success probability)} and Reward5: - \text{ave}{\it (energy)}.  The reward settings giving more weights to the worst-case problem instances such as Reward1 and Reward3 produce success probabilities more narrowly distributed. The averaging type of reward settings such as Reward2 and Reward4 yield a larger mean value of success probability but the distribution is much broader compared to Reward1 and Reward3. 
The energy type reward produces relatively lower success probability overall.}
\label{fig:distribution}
\end{figure}

\paragraph{Reward settings.}

To perform the AQC Hamiltonian schedule configuration, we implement the soft actor-critic method, a state-of-the-art RL algorithm dealing with continuous action control~\cite{haarnoja2018soft}.
We have tested several different ways of reward settings to connect SAC with AQC, including
Reward1:  {\it $\min$($\log$(success probability))} which denotes the minimum of the logarithm of success probabilities,
 Reward2: \text{ave}{\it ($\log$(success probability))} which denotes the average of the logarithm of success probabilities,
 Reward3: {\it $\min$(success probability)} which denotes the minimum of success probabilities,
 Reward4: \text{ave}{\it (success probability)} which denotes the average success probability, and
 Reward5: - \text{ave}{\it (energy)} which denotes the opposite of the average energy. The results with the five different reward settings are shown in Figure~\ref{fig:distribution}. 
 With Reward1 setting, the configured AQC algorithm  produces a relatively high  mean value of success probability $\sim 0.145$ which is larger than $P_\mathrm{th}$, and the success probability distribution is narrowly distributed around the mean value. 
Reward2 setting yields a larger mean value of success probability, but its distribution is too broad---there is a large number of  problem instances having success probability below $P_{\rm th}$.  
And the Reward3 setting leads to a similar performance as Reward1, with a mean success probability slightly smaller.
The case with Reward4 setting also produces a high mean value of success probability with a broader distribution which is similar to the performance of Reward2 setting. 
With Reward5 setting,  although the success probability has a narrow distribution, the mean value is much lower than other reward settings. 
Through these numerical tests, we conclude that it needs a proper type of reward signal to the SAC agent in the AQC configuration tasks to gain high success probability for all problem instances.  
It is a crucial prerequisite in the RL-based AQC configuration design scheme for the hard prime factorization instances. Conducting training processes under different reward settings leads to distinctive information flowing in the reinforcement learning scheme. The reward setting taking the success probability of the hardest instances (such as Reward1 and Reward3) provides a better guidance for the reinforcement learning method to configure the AQC algorithm. We choose the Reward1 setting in the following. 



\paragraph{Training process.}


\begin{figure}[htp]
\includegraphics[width=.48\textwidth]{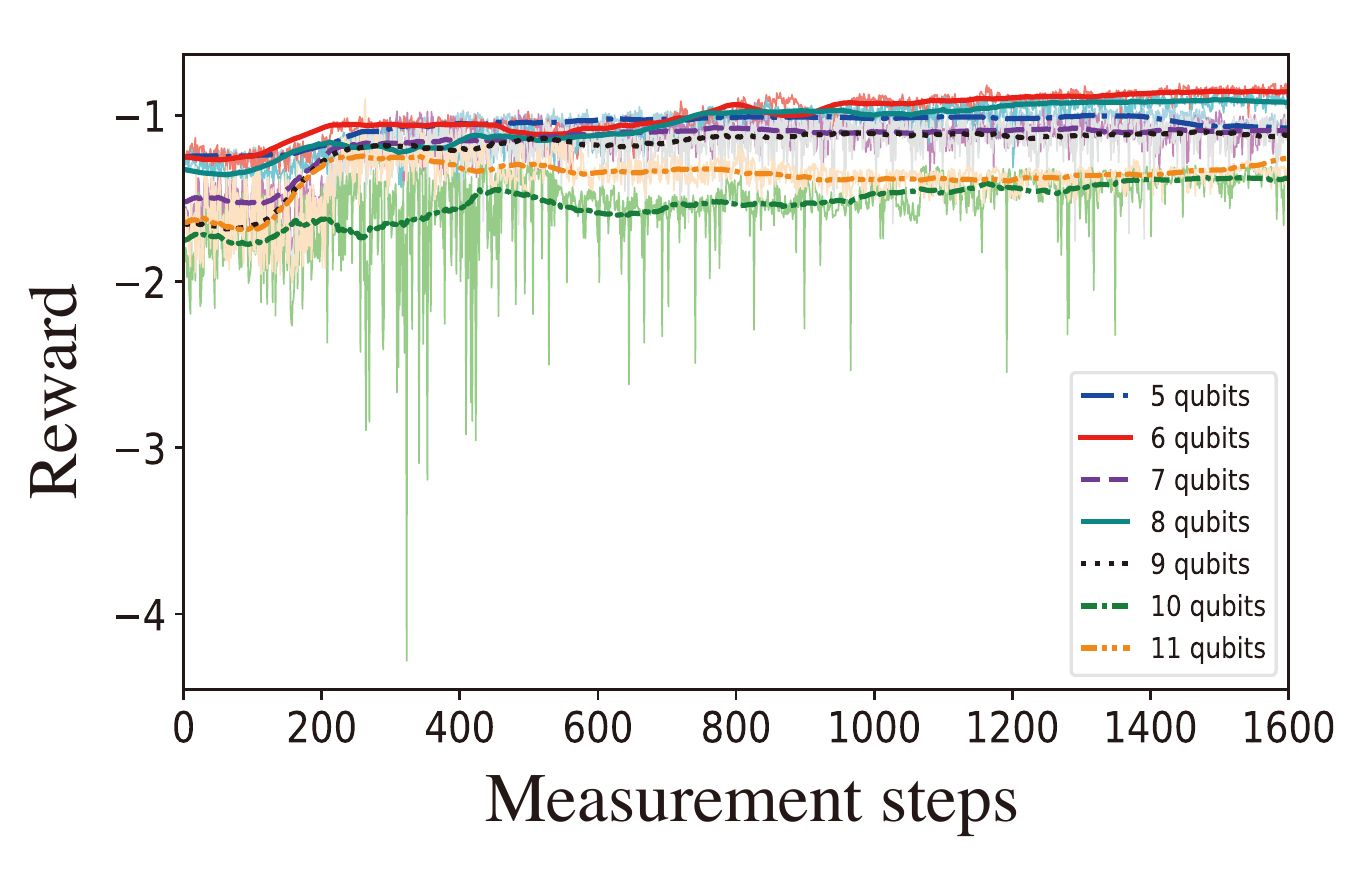}

\caption{{\textbf{The training process of SAC configuration on AQC-based prime factorization in the different system sizes with the qubit number of 5, 6, 7, 8, 9, 10, 11.} We record the reward of {\it $\min$($\log$(success probability))}  which corresponds to the performance of the hardest factorizing instance in each system size during the training. We observe that the SAC improves the performance of the hardest instances and all the cases of rewards converge within 1600 measurement steps. 
}
}
\label{fig:training}
\end{figure}

We apply SAC configuration on AQC-based prime factorization for a range of composite numbers from $N = 49$ to $N = 633$, whose Hamiltonian encoding has qubit numbers $n = 5, 6, 7, 8, 9,10,11$. We choose the AQC evolution time $T(n)$ according to the time when the average success probability of all the instances reaches the threshold ($P_\mathrm{th} = 0.1 $) in using the quadratic Hamiltonian schedule. From Figure~\ref{fig:training}, it is evident that the reward in SAC converges for all qubit numbers within 1600 measurement steps. Taking the RL-designed Hamiltonian schedule for AQC, the quantum factorization for all hard instances has a satisfactory success probability, roughly uniformly distributed. This results from our careful reward choice of using the minimum of the logarithm of success probability in training.

The slowing down problem caused by hard instances is thus rescued with our RL-configured AQC.
We remark here that if a higher success probability ($P^\star$) is upon request, this can be achieved by simply repeating AQC multiple ($M$) times, with $M=[{\log (1-P^\star)]/[\log (1-P_\mathrm{th})}]$~\cite{albash2018adiabatic}.
This relies on the fact that all the problem instances have a success probability above $P_\mathrm{th}$. This repeating protocol does not necessarily work if only the average success probability reaches $P_\mathrm{th}$.

\paragraph{Configured Hamiltonian schedules.}

\begin{figure}[htp]
\includegraphics[width=.48\textwidth]{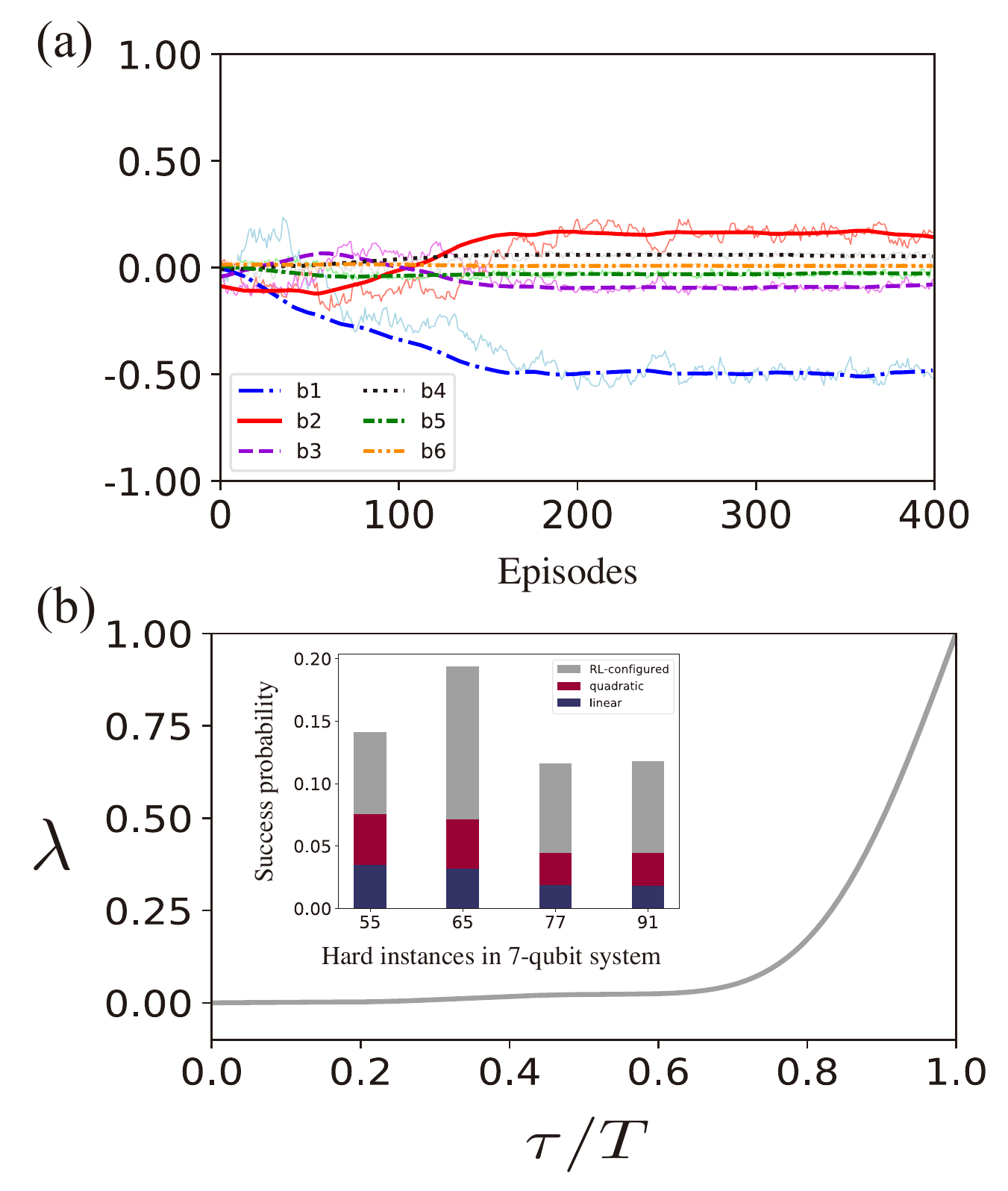}
\caption{\textbf{The evolution of $\mathbf{b}$ during the SAC training process in the 7-qubit system.} (a), the $\mathbf{b}$ values converge after 200 training episodes. For each episode, there are 4 measurement steps (see more details in the Supplementary Material). (b), the representative RL-configured Hamiltonian schedule and the performance comparison of different schedules. We observe the RL-configured schedule has a large flatten region at the beginning. The corresponding performance of different schedules are shown in the subplot. The success probabilities of RL-configured schedule which is trained from a linear one are dramatically improved.}
\label{fig:pathtype}
\end{figure}

We investigate the Hamiltonian schedule during the  RL learning process for the prime factorization instances encoded by 7 qubits. During the training of SAC, the reward signal converges within 1600 measurement steps (400 episodes with 4 measurements per episode) as shown in Figure~\ref{fig:training}.
 We observe that  the Hamiltonian schedule parameters $\mathbf{b}$  converge to a flattened plateau as shown in Figure~\ref{fig:pathtype}(a).
The RL-configured Hamiltonian schedule and the performances of different schedules on success probability are shown in Figure~\ref{fig:pathtype}(b). 
{One characteristic feature of RL-configured schedule is that it flattens at the beginning , and can be approximated by a power-law $\lambda \sim (\tau/T)^7$.}

{The RL-configured Hamiltonian schedule strongly deviates from the un-configured linear or quadratic schedule.}
{The AQC performance based on the RL-configured Hamiltonian schedule is shown in the inset} of Figure~\ref{fig:pathtype}(b). The factorized integers are $N = 55, 66,77,91,$ which are hard instances with un-configured AQC. {It is evident that the RL-configured Hamiltonian schedules exhibit substantially improved success probability---the success probability has doubled for the hard instances.}

\begin{figure}[htp]
\includegraphics[width=.48\textwidth]{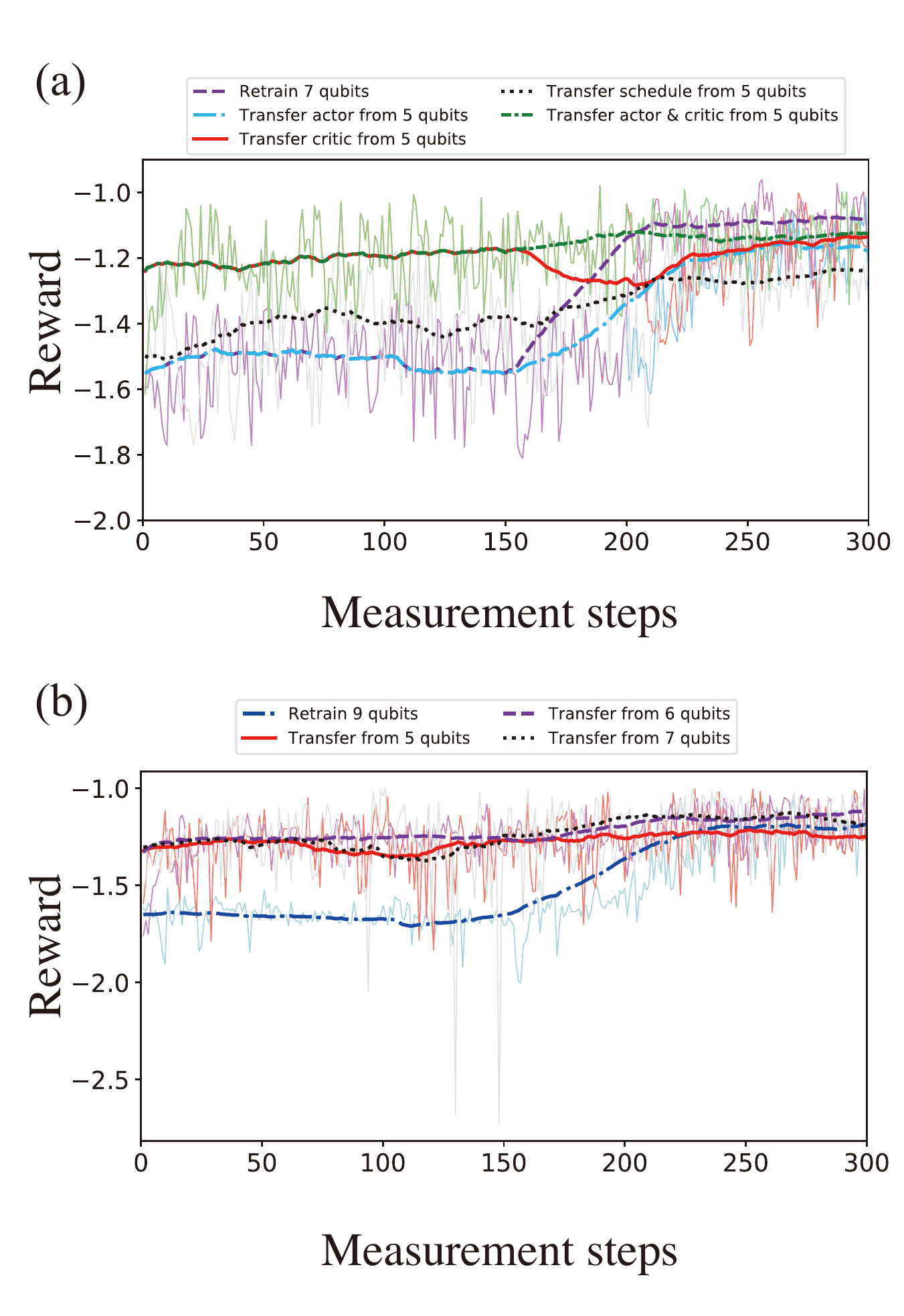}
\caption{\textbf{Transferring methods in hard instance learning.}  
The reward evolution during the learning process is shown in this plot. 
(a), Comparison of different transfer protocols on 7-qubit system. We transfer the network data trained on 5-qubit AQC to 7-qubit here. 
The learning process  of a direct retraining on 7-qubit system is shown as a baseline  (`purple dashed' line). 
The different protocols of transferring actor weights, critic weights,  all network weights, and Hamiltonian schedule are presented by `blue dashed dotted', `red solid', `green dash dotted' and  `black dotted' lines, respectively, in this plot. (b), Training curves of transferring the weights of actor \& critic networks trained on 5, 6, 7-qubit systems to 9-qubit system. 
All the transfer learning processes reach to an almost optimal reward within only a few measurement steps, much less than the direct training.  
}
\label{fig:transfer}
\end{figure}

\paragraph{Transferability.}

To further reduce the measurement cost, we investigate the transferability of our scheme. 
As the weights of learning networks are recorded during the training process, we test the different weight transferring methods, including Case ~\Rom{1}: transferring the recorded weights of actor networks, Case ~\Rom{2}: transferring the recorded weights of critic networks, Case ~\Rom{3}: transferring both recorded weights of actor \& critic networks. We also test the case of directly transferring the previous trained schedule as the initial ansatz in the new system size. 
We perform numerical tests by transferring from 5-qubit AQC to 7-qubit.  
The results are shown in Figure~\ref{fig:transfer}(a).

We find that the transfer protocol of Case III has a similar performance as Case II at the starting 150 measurement steps. After that, it is evident that Case III is more stable---it converges to optimal reward in a more systematic manner. 
The comparison between the transfer protocol of Case I and a direct training on 7-qubit system is similar. Their collected rewards are very close to each other at the starting 150 measurement steps, with tiny difference unnoticeable in the plot. After that, the Case I transfer protocol is even worse than the direct retraining. 
Through these numerical tests, we conclude that the protocol of transferring actor \& critic weights together outperforms other schemes in our AQC configuration task.  

To test the robustness of the transfer learning protocol of taking all actor \& critic weights,  we further apply this protocol to  a system of 9 qubits. The results are shown in Fig.~\ref{fig:transfer}(b). We transfer the weights trained on AQC with 5, 6, and 7 qubits to the 9-qubit system. These training processes taking the transfer data all have a faster convergence speed than a direct retraining on 9-qubit system, confirming the robustness of the transfer learning protocol of taking all weights. 
The applicability of this type of transfer learning for AQC configuration from smaller number of qubits to larger sizes also implies the scalability of our scheme.

\section{Conclusion}

We develop a reinforcement learning based scheme for adiabatic quantum algorithm configuration directly targeting the hard prime factorization instances.
This is achieved by  taking the minimum of the logarithm of the success probability of the AQC on factorization as the RL reward.
By implementing the SAC algorithm, we find the learning process converges within only a few hundred measurements.
Through numerical tests, we have shown that our developed AQC algorithm configuration scheme has fair transferability---the learning transferred from smaller qubit number systems to larger qubit numbers converges significantly faster than learning from scratch. This implies our AQC algorithm configuration scheme is potentially quite scalable.

\section{Methods}

\paragraph{Adiabatic Algorithm Design as Markov Decision Process}

Aiming at an automated architecture for optimal design of quantum adiabatic factorization, we use a Markov decision process (MDP)~\cite{richard1998intro} for the configuration of the $\mathbf{b}$-parametrized Hamiltonian schedule. In MDP, we update the vector $\mathbf{b}$ according to a stochastic policy $\pi$, which generates a sequence $\mathbf{b}_{t+1}= \mathbf{b}_{t} + {\bf a}_t$, with ${\bf a}_t$ random variables drawn according to $\pi$.
In our study, we consider a MDP with a finite-depth ($L_\mathbf{M}$).
The  optimal quantum adiabatic algorithm design is then converted to searching for a MDP policy converging to a  $\mathbf{b}$-vector that maximizes the success probability of AQC.

Quantitatively, the optimal policy is defined to be one that maximizes an objective of long-term reward $\mathbf{E} _\pi[\sum_t(\gamma^tr_t)]$, with $\gamma\in(0,1]$ a discount factor, and $r_t$ a reward at $t$-th step. The reward is defined by the success probability of AQC taking the configuration $\mathbf{b}_t$.

We compare the performances of different reward settings and choose min(log({\it success probability})) as the reward. This choice automatically gives extra weight to the most difficult problem instances having low success probability.

There are multiple RL protocols to specify the MDP policy.
Here we focus on the entropy-regularized soft actor-critic (SAC) algorithm~\cite{haarnoja2018soft}. For SAC, the objective return is modified to the form of $\mathbf{E}_\pi[\sum_t(\gamma^tr_t+\alpha \mathcal{H}(\pi(\cdot|s_t)))]$ where $\mathcal{H}$ is the causal policy entropy and $\alpha$ determines its relative weight.
The modification balances the trade-off between exploration and exploitation. The state of RL-agent is specified to be $\mathbf{s}_t = \{\operatorname{DoubleOneHotEncoder}(t), \mathbf{b}_t\}$, with more details of problem setting discussed in the Supplementary Material. At each iteration, the agent chooses an action $\mathbf{a}_t$ by sampling from the actor network, and acts (adds) to $\mathbf{b}_t$. We assume $\mathbf{b}_t \in [-1, 1]$, and set the action element $a_m\in[-0.01,0.01]$.  In SAC training, the actor is supervised by critic, who processes reward signal and gets updated according to the temporal-difference error using Polyak averaging~\cite{polyak} with a target to stabilize training.
The whole architecture for our quantum algorithm configuration is illustrated in Figure~\ref{fig:architecture}. More details are provided in the Supplementary Material.

\paragraph{The pseudo code of reinforcement learning process}
\mbox{}

\begin{algorithm}[H]

  \caption{SAC-Configured AQC}
  \label{alg:RQAC}
  \def\Loss{\operatorname{\mathcal{L}oss}}
  \begin{algorithmic}
  \REQUIRE
    $\mathbf{X}$: prime factorization instances;
    $T$: evolution time;
    $\mathbf{\theta}$: initialized policy parameters;
    $\mathbf{\phi},\hat{\mathbf{\phi}}$: initialized critic and target parameters;
    $\mathbf{b}_0$: initialized schedule;
    $\Delta t_w$: measure frequency;
    $P_{\mathrm{th}}$: success probability threshold
  \ENSURE
    configured schedule $b_1, b_2, \cdots, b_C$
  \STATE $\mathbf{H}_{\text{p}} \longleftarrow \operatorname{QUBOEncoding}(\mathbf{X})$
  \STATE $\mathbf{X_{\text{hard}} \longleftarrow} \operatorname{SELECT}(\mathbf{X}, P_{\mathrm{th}})$
  \STATE $\mathbf{b} \leftarrow \mathbf{b}_0$
  \FOR{each episode}
    \FOR{each step $t$ in episode length}
      \STATE $\mathbf{s}_t \leftarrow [\operatorname{DoubleOneHot}(t), \mathbf{b}_t]$
      \STATE Sample $\mathbf{a}_t \sim \mathbf{\pi}_\theta(\mathbf{s}_t)$
      \STATE $\mathbf{b}_{t+1} \leftarrow \mathbf{b}_t + \mathbf{a}_t$
      \WHILE{$\operatorname{MonotonicConstraint}(\mathbf{b}_{t+1})$ is false}
        \STATE Resample $\mathbf{a}_t \sim \mathbf{\pi}_\theta(\mathbf{s}_t)$
        \STATE $\mathbf{b}_{t+1} \leftarrow \mathbf{b}_t + \mathbf{a}_t$
      \ENDWHILE
      \STATE $\mathbf{s}_{t+1} \leftarrow [\operatorname{DoubleOneHot}(t+1), \mathbf{b}_{t+1}]$
      \IF{$t \bmod \Delta t_w = 0$}
        \STATE $r_t \leftarrow \operatorname{QuantumEvolution}(\mathbf{b}_{t+1},T,\mathbf{X}_{\text{hard}})$
      \ELSE
        \STATE $r_t \leftarrow 0$
      \ENDIF
    \ENDFOR
    \STATE $\text{Buffer} \leftarrow (\mathbf{s}_t,\mathbf{a}_t,r_t,\mathbf{s}_{t+1})$
    \FOR{each gradient step}
      \STATE $\bm{\theta} \leftarrow \bm{\theta}-\nabla_{\bm{\phi}} \Loss(\bm{\phi})$
      \STATE $\bm{\phi} \leftarrow \bm{\phi}-\nabla_{\bm{\phi}} \Loss(\bm{\phi},\hat{\bm{\phi}})$
      \STATE $\hat{\bm{\phi}} \leftarrow \eta\bm{\phi} + (1-\eta)\hat{\bm{\phi}}$
    \ENDFOR
  \ENDFOR

  \end{algorithmic}
\end{algorithm}

\section{Acknowledgements}
J.L. acknowledges helpful discussion with Xiangdong Zeng. This work is supported by National Natural Science Foundation of China (Grant No. 11774067 and 11934002), National Program on Key Basic Research Project of China (Grant No. 2017YFA0304204), and Shanghai Municipal Science and Technology Major Project (Grant No. 2019SHZDZX01), Shanghai Science Foundation (Grant No. 19ZR1471500).

\section{Author contributions}
X.L. conceived the main idea; J.L. and Z.F.Z. developed the
methods and performed numerical calculations. All authors cotributed to analyzing the numerical test and to the preparation of manuscript.

\section{Competing interests}
The authors declare no competing interests.


\bibliographystyle{naturemag}
\bibliography{references}

\clearpage

\onecolumngrid

\ctexset{
  section = {
    format = \LARGE\centering,
    afterskip = 3ex,
  },
  subsection/number = \Alph{subsection},
  appendix/numbering = false,
}

\appendix
\section{Supplementary Material}

In this supplementary material, we present details of the Hamiltonian encoding, the environment setups, and the hyperparameters used in our RL-based AQC algorithm configuration.

\subsection{Encoding Protocol: Quadratic Unconstrained Binary Optimization (QUBO)}

Here we provide the details of mapping the factorization problem to the quadratic unconstrained binary optimization (QUBO) problem. The problem of prime factorization is to factorize a given integer $N$ into two integers $p$, and $q$, i.e., $N\to p\times q$.

For the first step, we divide the multiplication table (such as Tab.~\ref{tab1}) and calculate the total qubit number needed to factorize $N$. We define the bit length $L_p = \lfloor \log_2 p \rfloor$, $L_q = \left \lfloor \log_2 q \right \rfloor$, $L_N = \left \lfloor \log_2 N \right \rfloor$ as well as the bit string of $p:(p_{L_p}=1,p_{L_p-1},\ldots,p_1,p_0=1)_2$ and $q:(q_{L_q}=1,q_{L_q-1},\ldots,q_1,q_0=1)_2$, where $\left \lfloor a \right \rfloor$ denotes the largest integer not larger than $a$ and assuming $p\le q$ without loss of generality. We divide the multiplication table into $B_K$ blocks with the width $W$, then $B_K = \left \lfloor (L_p+L_q)/W \right \rfloor$. We define $\rho_i$ as the sum of the product terms in the block $i (i\ge 1)$ :
\begin{equation}
\rho_i = \sum_{j=1+(i-1)\cdot W}^{i\cdot W}2^{j-1-(i-1)\cdot W}\sum_{\substack{m+n=j \\ m\in[0,L_p] \\ n\in[0,L_q]}} p_m*q_n.
\end{equation}.


We calculate the maximum value of carry variables and product terms in each block $i (i\ge 1)$ :

\begin{equation}
\mathscr{B}_i = \mathscr{C}_{i-1}+\mathop{max} \rho_i,
\end{equation}

where $\mathscr{C}_0=0 $. To obtain $\mathscr{C}_i (i\ge 1)$, we define $\mathscr{M}_i=\lfloor \mathscr{B}_i/2^{W} \rfloor$ and the number of carry variables $c_i$ into block $i+1$

\begin{equation}
c_i = \left\{
\begin{aligned}
&0,
&\text{if\ } \mathscr{M}_i= 0;\\
&\lfloor \log_2(\mathscr{M}_i) \rfloor+1,
&\text{otherwise},
\end{aligned}
\right.
\end{equation}
then $\mathscr{C}_i=2^{c_i}-1$. And we calculate $\chi_i = \sum_{k=1}^{i-1}c_k$, the total number of carry variables before (including) block $i$.

In the following, the total number of carry bits and summation terms in the cost function are defined as  $T_C$ and $P_N$, respectively. If $L_N-B_K\cdot W > 1$, then $T_C =\sum_{i=1}^{B_K} c_i$ and $P_N = B_K+1$. Otherwise, $T_C = \sum_{i=1}^{B_K-1}c_i,P_N = B_K$. The total auxiliary variable number $T_A$ in reducing the higher order coupling terms is $(L_p-1)\cdot(L_q-1)$. So the total qubit number required is
\begin{equation}
T_Q = (L_p-1)+(L_q-1)+T_C+T_A.
\end{equation}

Then we turn to the second step to construct the cost function .
We define $K_i$ as the sum of carry variable $\tilde{C}$ into block $i$:
\begin{equation}
K_i = \sum_{j=1}^{c_{i-1}}2^{j-1}\tilde{C}_{\chi_{i-1}+j},
\end{equation}
and $F_i$ as the sum of the carry variable $\tilde{C}$ into block $i+1$:

\begin{equation}
F_i = \sum_{j=1}^{c_i}2^{W+j-1}\tilde{C}_{\chi_{i}+j},
\end{equation}
as well as $V_i$ the target value of the block $i$ which can be read out from the binary presentation of $N$ directly.

We can write down the total cost function as:
\begin{equation}
f_\text{cost} = \sum_{i=1}^{P_N}(\rho_i+K_i-F_i-V_i)^2,
\label{cost}
\end{equation}
which can be expanded and simplified using $x^2 = x$  for $x=0,1$.

Then we turn to the third step to introduce auxiliary variables to reduce $k$-bit coupling terms ($k\ge3$) using

\begin{equation}
\begin{cases}
x_1x_2x_3 = \min_{x_4}(x_4x_3+2(x_1x_2-2x_1x_4-2x_2x_4+3x_4); \\
-x_1x_2x_3 = \min_{x_4}(-x_4x_3+2(x_1x_2-2x_1x_4-2x_2x_4+3x_4),
\label{reduce}
\end{cases}
\end{equation}
by which the cost function is reduced to a quadratic form. The problem of prime factorization is then converted to the quadratic unconstrained binary optimization (QUBO) problem.

\begin{table}
\begin{center}
\def\EMPTY#1{\multicolumn{1}{c}{#1}}
\begin{tabular}{cc|ccc|ccc|c}
  \hline \\[-1.8ex]
  & \EMPTY{$2^7$} & $2^6$ & $2^5$ & \EMPTY{$2^4$} & $2^3$ & $2^2$ & \EMPTY{$2^1$} & $2^0$ \\[0.6ex]
  \hline\hline & \EMPTY{} & & & \EMPTY{} & & & \EMPTY{} & \\[-1.8ex]
  $p$ & \EMPTY{} & & & \EMPTY{} & 1 & $p_2$ & \EMPTY{$p_1$} & 1 \\
  $q$ & \EMPTY{} & & & \EMPTY{} & 1 & $q_2$ & \EMPTY{$q_1$} & 1 \\[0.6ex]
  \hline & \EMPTY{} & & & \EMPTY{} & & & & \\[-1.8ex]
  & \EMPTY{} & & & \EMPTY{} & 1 & $p_2$ & $p_1$ & 1 \\
  & \EMPTY{} & & & $q_1$ & $p_2q_1$ & $p_1q_1$ & $q_1$ & \\
  & \EMPTY{} & & $q_2$ & $p_2q_2$ & $p_1q_2$ & $q_2$ & & \\
  & & 1 & $p_2$ & $p_1$ & 1 & & \\
  carries & & & $\tilde{C}_2$ & $\tilde{C}_1$ & & &  & \\[0.6ex]
  \hline & & & & & & & & \\[-1.8ex]
  $p\times q=143$ & 1 & 0 & 0 & 0 & 1 & 1 & 1 & 1 \\[0.6ex]
  \hline
\end{tabular}

\caption{Multiplication table for $143=11\times13$. The $p$ and $q$ rows are the binary representation of the prime factor and $\tilde{C}_i$ in carries row is the carry bit. The bottom row is the binary representation of the number $N = 143$ }
\label{tab1}
\end{center}
\end{table}

As the last step, we replace $x_i$ as $(1-\sigma_{z_i})/2$ where $\sigma_{z_i}$ is the Pauli $Z$ matrix and convert the cost function into the Ising type Hamiltonian $H$ which would be suitable for AQC.

As an illustration, we give a multiplication table for $143=11\times13$ in Tab.\ref{tab1}.

The resulting equations derived from the two blocks are :

\begin{align}
(1+p_2q_1+p_1q_2+1)\times2^2+(p_2+p_1q_1+q_2)\times 2+(p_1+q_1) &= \tilde{C}_2\times 2^4+\tilde{C}_1\times2^3+(111)_2\notag \\
& = 16\tilde{C}_2 +8\tilde{C}_1 +7.
\label{eq:bk}
\end{align}

\begin{align}
1\times 2^2+(q_2+p_2+\tilde{C}_2)\times 2+(q_1+p_2q_2+p_1+\tilde{C}_1)&= (1000)_2\notag \\
& = 8.
\label{eq:bk2}
\end{align}

The corresponding cost functions are:
\begin{equation}
f_1 = (4p_2q_1+4p_1q_2+2p_1q_1+2p_2+2q_2+p_1+q_1-16\tilde{C}_2-8\tilde{C}_1+1)^2,
\end{equation}
\begin{equation}
f_2 = (p_2q_2+2p_2+p_1+2q_2+q_1+2\tilde{C}_2+\tilde{C}_1-4)^2.
\end{equation}
The high order $k(k\ge3)$ terms in the cost function can be reduced using  Eq.~\ref{reduce}. As mentioned before, with the variable replacement in the last step, we can get the Ising type Hamiltonian.

\subsection{DoubleOneHot Environment}

The episode length, quantum schedule and time-step in the episode are denoted by $L$, ${\mathbf{b}}\in \mathcal R^6$ and $t\in [1,L]$, respectively.

\textbf{State.} State $s_t$ contains a sequential encoder, in which we use two one-hot encoders with 10-bits respectively, representing 100 steps at large with fixed 20 dimensions. Hence, this sequential encoder turns episode length $L$ into a one-hot vector with 20 dimensions and 6 quantum schedule is denoted as ${\mathbf{b}}\in \mathcal R^6$.

\textbf{Action.} The learned policy maps state into action denoted as ${\mathbf{a}}\in \mathcal R^6$.
At each step $t$, we define action ${\mathbf{a}}_t \in \mathcal R^6$ as how far should the agent reach based on current schedule ${\mathbf{b}}_t \in \mathcal R^6$and current
step encoded by two one-hot encoders denoted as $\text{double-one-hot}(t)$.

\textbf{Reward.} Measure the action ${\mathbf{a}}_t$ sampled from policy $\pi_\theta$ if it satisfies the monotonic constraint enforced on the evolution path. And return the minimum logarithm of success probability as a reward signal.

\textbf{DoubleOneHotEnv.} The problem setting can be abbreviated as follows:
\begin{align*}
    \mathbf{s}_t&=[\text{double-one-hot}(t),{\mathbf{b}}_t]\in \mathcal R^{26}, t\in \{1,2,\cdots,100\}\\
    \pi & : {\mathbf{a}}_t\sim \pi_\theta(\cdot|s_t) \\
    r_t &= \text{Measure}({\mathbf{a}}_t + {\mathbf{b}}_t \triangleq {\mathbf{b}}_{t+1})\\
   \mathbf{s}_{t+1} &= [\text{double-one-hot}(t+1),{\mathbf{b}}_{t+1}]\in \mathcal R^{26}
\end{align*}

\subsection{Hyperparameters}

Hyperparameters used in the SAC-configured AQC Algorithm are listed below, which are categorized into \emph{Env}, \emph{Learner} and \emph{Network}.
In terms of \emph{Env} hyperparameters, we specify the quantum schedule space and action stride at first and then formulate state and Markov decision process as known environment setting. In subsequence, the exploration space is mainly determined by episode length. The reward signal is measured once after $n$ steps, and amplified by reward scale. Finally, other direct hyperparameters about \emph{Learner} and \emph{Network} are listed as follows.

\begin{table}[H]
\renewcommand{\arraystretch}{1.1}
\centering
\caption{SAC-configured AQC Parameters}
\label{table:parameters}
\vspace{1mm}
\begin{tabular}{l l| l }
\toprule
\multicolumn{2}{l|}{Parameter} &  Value\\
\midrule
\multicolumn{2}{l|}{\textit{Env}}& \\
& action stride & $[-0.01, 0.01]$                \\
& schedule space & $[-1.0,1.0]$                    \\
& environment setting & DoubleOneHotEnv         \\
& number of episodes & 1000                  \\
& episode length & 40                         \\
& reward scale  & 5                         \\
& measure every n steps & 10                     \\
& reward type & min(log(success probability))                    \\

\midrule
\multicolumn{2}{l|}{\textit{Learner}}& \\
& optimizer &   Adam                            \\
& learning rate & $3 \cdot 10^{-4}$             \\
& discount ($\gamma$) &  0.9                   \\
& alpha & 0.02                                \\
& polyak & 0.995                              \\
& target update interval & 2                    \\
& gradient steps & 2                          \\
\midrule
\multicolumn{2}{l|}{\textit{Network}}& \\
& number of hidden layers & 2                   \\
& actor hidden layer size & 256                         \\
& critic hidden layer size & 512                         \\
& batch size & 128                            \\
& actor maximum std & 1                            \\
& actor minimum std & $-10$                           \\
& random steps & 0                         \\
& buffer size & $10^6$                          \\

\bottomrule
\end{tabular}
\end{table}

\end{document}